\documentclass[11pt,preprint,epsfig]{aastex}
\usepackage{emulateapj5,epsfig}

\slugcomment{Accepted for publication in the ApJ}
\shortauthors{Kronberg et. al.}
\begin{document}
\title{Magnetic Energy of the Intergalactic Medium from Galactic Black Holes}

\author{P. P. Kronberg\altaffilmark{1}, Q. W. Dufton\altaffilmark{1},
H. Li\altaffilmark{2} and S. A. Colgate\altaffilmark{3}}

\altaffiltext{1}{Department of Physics, University of
Toronto, 60 St. George St., Toronto M5S 1A7, Canada;
kronberg@physics.utoronto.ca;
dufton@physics.utoronto.ca }
\altaffiltext{2}{Applied Physics Division,  MS B288,
Los Alamos National Laboratory, Los Alamos, NM 87545;
hli@lanl.gov }
\altaffiltext{3}{Theoretical Astrophysics, T-6, MS B288,
Los Alamos National Laboratory, Los Alamos, NM 87545;
colgate@lanl.gov }

\begin{abstract}

We present a quantitative analysis of two radio source samples 
having opposite extremes of ambient gas density that
leads to important new conclusions about the magnetic energy 
in the intergalactic medium (IGM). 
We analyze here (i) a new, large sample of well imaged ``giant'' 
extragalactic radio sources that are found in rarified IGM 
environments and (ii) at the other extreme, radio galaxies situated 
in the densest known IGM environments, within 150 kpc of 
rich cluster cores. We find that sources in the former sample
contain magnetic energies $E_{\rm B} \sim 10^{60-61}$ ergs
and could be viewed as important ``calorimeters'' of the 
{\it minimum} energy a black hole (BH) accretion disk system
injects into the IGM. In contrast to the radiation energy
released by BH accretion, most of the magnetic energy is ``trapped'' 
initially in a volume, up to $\sim 10^{73}$ cm$^{3}$, around the 
host galaxy. But since these large, Mpc scale radio 
lobes are still overpressured after the AGN phase, their 
subsequent expansion and diffusion will magnetize a large 
fraction of the entire IGM. This suggests that 
the energy stored in intergalactic magnetic fields will have 
a major, as yet underestimated effect on the evolution of 
subsequently forming galaxies. Comparison with the second, 
cluster core-embedded sample shows that the minimum magnetic 
energy $E_{\rm B}$ can be a strongly variable fraction of the inferred
accretion energy $E_{\rm acc}$, and that it depends on the ambient IGM 
environment. Cluster embedded AGNs inject significant energy 
as $PdV$ work on the thermal ICM gas, and their magnetic energy, 
even ignoring the 
contribution from stellar and starburst outflows, is sufficient to 
account for that recently found beyond the inner cores of galaxy 
clusters. We discuss the various energy loss processes as these 
magnetized CR clouds (lobes) undergo their enormous expansion 
into the IGM. 
We conclude that the aggregate IGM magnetic energy derived 
purely from galactic black holes since the first epoch of significant 
galaxy BH formation is sufficiently large that it will have an 
important influence on the process of both galaxy and visible structure formation on scales up to $\sim 1$Mpc.

\end{abstract}


\section{INTRODUCTION}

Advances in observational techniques have revealed the widespread
existence of magnetic fields in the Universe 
[see Kronberg (1994) for a review]. 
Important questions at this point are how strong are the magnetic
fields as derived from 
current observations, how widely are they distributed, where are they 
seeded, where and how are they amplified, and how much do they 
contribute to the energy budget of the intergalactic medium? 
It was pointed out over 30 years ago (Burbidge 1956; 1958) that a 
single galaxy releases a very large magnetic energy $-$ up 
to $\sim 10^{61}$ ergs, and that gravitational energy is the only 
feasible source (Hoyle et al. 1964, Burbidge \& Burbidge 1965). 
 
An excellent review by Begelman, Blandford, \& Rees (1984) 
made a strong case for accretion on the central supermassive black holes 
as the energy engine for the powerful radio sources. Further progress 
has been made toward answering some of these questions
(e.g., Bridle \& Perley 1984; Wan, Daly, Guerra 2000), and 
this paper presents a new analysis which focuses on AGN-fed, 
extended radio sources and their immediate 
intergalactic environment. The analysis supports recent arguments 
(Colgate \& Li 1999; 2000; Colgate, Li, \& Pariev 2001) that 
a strong feedback effect might exist on the
dynamics of the intergalactic medium (IGM) from the formation of 
supermassive black holes. A significant fraction of the energy 
released during
these formation events could have been directly converted 
into magnetic field energy and magnetic flux, which are injected into 
extragalactic space.

The fact that extragalactic radio sources are seen in synchrotron 
radiation enables an approximate calculation of the minimum energy 
contained in the sources' magnetic fields and relativistic particles. 
The jet-lobe morphology and commonly high polarization degree 
confirm, respectively, that the energizing source is at the host
galaxy/quasar nucleus, and that the largest field ordering scales are 
comparable to, or greater than a galactic dimension. 
This latter fact sets constraints on the magnetic field 
generation process. 

Recent Faraday rotation (RM) measurements, in combination with X-ray 
bremsstrahlung data have provided complementary probes of the strength 
and structure of magnetic fields {\it outside} of radio sources, but 
within the wider ambient hot {\it thermal} gas
in galaxy clusters (cf. Kim et al. 1989; 1990; Taylor \& Perley 1993; 
Feretti et al. 1995; Eilek 1999; Clarke, Kronberg, \& 
B\"{o}hringer 2001). These studies, which used polarized 
synchrotron-emitting sources inside and/or behind clusters, have 
shown (Taylor \& Perley 1993) that the central density-enhanced 
``cooling flow'' regions ($r \leq 150$kpc) have magnetic fields up to
$\sim 40 \mu $G, with coherence scales up to $\sim 50 \,$kpc. 
The most recent and definitive RM --- X-ray probe of the wider 
cluster ICM ($r \leq 1$Mpc) beyond any cooling flow zones by 
Clarke et al. (2001) used a combination of background and
cluster-internal radio sources to show that much of a rich galaxy 
cluster's volume out to $r\,=\,500$kpc is magnetized to a level 
of $5 \mu $G or more, implying a total cluster ICM magnetic energy 
of $E_{B} \simeq 1.5 \times 10^{61}({\frac{r}{500\ {\rm kpc}}})^{3}
(\frac{B} {5\ \mu {\rm G}})^{2}$\,$h_{75}^{-2}\ \rm ergs$. 
Given that cooling flows, to the extent they occur, are 
a later phenomenon of a cluster's history, this result shows 
that galaxy clusters have been significantly, and previously, 
magnetized by processes other than cooling flows. We will offer an 
explanation for this important new result. 

The important role of magnetic fields in clusters is also 
clearly demonstrated by the interaction between radio galaxies 
and their surrounding ICM, as revealed 
by recent Chandra observations. Large cavities in the X-ray emitting gas 
correspond strikingly with the radio-bright lobes of the 
radio galaxies in Hydra A (McNamara et al. 2000) and in the Perseus 
cluster (Fabian et al. 2000). In both cases, an energy of at
least $\sim 10^{59}$ ergs is needed in order to displace the X-ray gas
via $PdV$ work. This energy is presumably supplied by the expanding
magnetized radio lobes, and hence originated in the galaxy's nucleus.

In this paper, we have analyzed $\sim 100$ powerful extragalactic 
radio galaxies.
We use them as indicators of the minimum net amount of magnetic 
energy, hence total energy that comes from a galactic black 
hole/accretion disk system. Since extended extragalactic radio sources 
really form a continuous distribution in many ways,
we have chosen two categories with contrasting extremes of 
{\it external} ambient IGM pressure. These are:

(i) sources with large projected linear size,
$\geq 670\,h_{75}^{-1}$\,kpc, most of which are likely to be in a 
rarefied IGM environment. They are referred to as ``giant'' 
sources, and we have compiled here a substantial list 
of $\sim 70$ such sources, largely from the Northern Hemisphere, 
that are currently known and well imaged. 

(ii) $\sim 30$ sources located in the {\it densest} known 
IGM environment --- within $\sim 150$ kpc of the cores of rich 
clusters. We refer to them as ``cluster sources'' in this
paper.

We argue that these giant sources can be used as the best 
``calorimeters'' for the {\it minimum} amount of magnetic energy 
that galactic black holes have injected into intergalactic space 
since the cosmological epoch at which galactic black holes 
began to form.

We will conclude by showing that energies and magnetic flux 
injected into intergalactic space support recent work by Colgate \& Li 
(1999; 2000) and Colgate, Li, \& Pariev (2001), who proposed a mechanism 
for extracting the very large accretion energy of the commonly 
occurring $\sim 10^{8}M_{\odot}$ galactic black holes
and releasing it directly into large intergalactic volumes. We briefly 
discuss some of the implications for galaxy formation, 
the physics of the IGM and cosmic magnetic fields.

\section{MAGNETIC ENERGY IN RADIO GALAXIES}
\label{sec:rl}

\subsection{Data Compilation}

In Table 1, we compile approximate estimates of the minimum total
energy $E_{min}^{tot}$ (magnetic fields plus relativistic particles) and 
the minimum magnetic field $B_{minE}$ for a sample of well$-$imaged 
extragalactic radio sources from the recent literature that have a 
projected largest linear size (LLS, from lobe to lobe) of 
$\sim 0.67\,h_{75}^{-1}$ Mpc or greater. 
The source volumes were estimated assuming a cylindrical source 
shape, where the length and radius of the cylinder are estimated
based on the projected dimension as measured from the lowest 
reasonable contours from the radio images. 
We used the angular size distances, 
$D_{A}=D_{L}/(1+z)^{2}$, $D_{L}$ being the luminosity distance.

These giant radio sources were primarily identified from three recent
compilations of source data and images: Nilsson (1997), 
Ishwara-Chandra \& Saikia (1999), and Schoenmakers et al. (2000a). Some additional 
sources in Table 1 are taken from other papers if they qualified as 
giants on our criterion. Spectral indices, flux densities and other 
observable parameters were either obtained from the above primary 
sources, or from other articles identified in the table. The images used to
derive the numbers in Table 1 were at frequencies between 0.15 
and 5\,GHz, and where possible we also inter-compared radio images at 
lower and higher frequencies within this range. In all such cases the 
results were found not to be significantly dependent 
on the frequency of the radio image used to calculate $E_{min}^{tot}$. 
The {\it minimum} total energy $E_{min}^{tot}$ within a 
synchrotron-emitting volume ($V$) containing relativistic particles 
and magnetic field can be expressed in terms of the measurables, 
luminosity and volume 
\begin{equation}
E_{min}^{tot} = \left(\frac{3}{4\pi}\right)^{3/7} {\cal C}^{4/7} 
(1+k)^{4/7} (\phi V)^{3/7} L^{4/7}~{\rm ergs},
\end{equation}
where we have approximately followed Pacholczyk (1970). Here,
$k$ is the
relativistic proton to electron energy ratio, $\phi$ the volume 
filling factor of the synchrotron emission, 
$L$ the integrated radio luminosity, calculated between fixed 
frequencies ($\nu _{1}, \nu _{2}$)\,=$(10^{7}, 10^{10}$ Hz) in the 
emitted frame and $z$-corrected to the emitted frame, and ${\cal C}$ 
is a slowly varying function of $\alpha \,, \nu _{1}, \nu _{2}$, where 
${\alpha}$ is defined by $S\, \propto \nu ^{\alpha}$, and $S$ is 
source's spectral flux density. Because the radiating volumes 
reveal a characteristic filamentary structure, we conservatively adopted 
a global effective filling factor $\phi \, = 0.1$. Note that the 
total and magnetic energies are only mildly sensitive to $\phi $. 
The total energy is minimized when the magnetic field associated with 
the synchrotron radiation has the value 
\begin{equation}
B_{minE} =({6\pi})^{2/7} 
(1+k)^{2/7}\,{\cal C}^{2/7}
(\phi V)^{-2/7} L^{2/7}~{\rm G},
\end{equation}
again following the terminology of Pacholczyk (1970). 
Our calculations of $E_{min}^{tot}$ and $B_{minE}$ assume $k=100$, 
which is close to the measured value for Galactic cosmic rays.
The estimated $E_{min}^{tot}$ and $B_{minE}$ will be 
reduced by a factor of $\sim 14$ and $\sim 4$, respectively, if
$k=0$ is used. Furthermore, using $\phi=1$ will 
increase $E_{min}^{tot}$ by a factor of $2.7$ but decrease
$B_{minE}$ by $2$. 

The integrated flux densities for each source were checked against
standard compilations of flux densities to test if the image used
contained a substantial fraction of the source's entire 
synchrotron luminosity. Where substantial differences 
were noted alternative images were selected. 
The source volume estimates (col. (6) of Table 1) were 
based on the most sensitive available image, 
and $H_0 =75, \Omega =1$. The volume, 
hence total energy and magnetic field estimates in Table 1 must be 
interpreted as only global, approximate estimates for any given
source, since an optimally precise calculation of $E_{min}^{tot}$ 
and $B_{minE}$ would require a detailed integration over 
an assumed 3-D emissivity distribution. This is not possible to do in 
a consistent way at present, given the inhomogeneity of the currently 
available sample. Our global estimate of the uncertainties are 
$\sim \pm 30 \%$ for $E_{min}^{tot}$. Such errors do 
not include systematics such as (a) the unknown projection angle of a 
given source into the plane of the sky, or (b) any undetected 
``halo flux'' --- both of which could systematically
increase our estimates of $E_{min}^{tot}$ by an unknown
but potentially large amount (e.g. the extended halo around M87).

Table 2 lists the same measured and calculated quantities as in Table
1, but is restricted to extended radio galaxies that are 
located within 150 kpc (projected) of the cores of rich galaxy
clusters (see Table 2 for references). These cluster sources include some 
very well-studied sources, for some of which there are not only detailed 
images of the synchrotron radio emission, but also of Faraday rotation 
and detailed recent X-ray images from the ROSAT and/or Chandra satellites. 
Such cases allow both the sources and their cluster environments 
to be independently probed; That is, we can compare the energy in the 
synchrotron-emitting cosmic ray gas magnetic fields with that of the 
ambient thermal ICM. Where possible, the cluster source 
morphologies were compared 
in detail with recent X-ray images. The latter show a striking effect, 
namely that the bremsstrahlung-emitting ambient intracluster gas 
gets displaced by 
the cosmic ray gas of the radio lobes. Recent examples are the 
depressions in the (projected) X-ray surface brightness that clearly 
coincide with the periphery of radio lobes of Hydra A 
(McNamara et al. 2000)
and the arcmin-scale lobes of Perseus A (3C84) (Fabian et al. 2000). 
In addition, there have been a half dozen or so detections of
X-rays from knots/jets (e.g., Wilson, Young, \& Shopbell 2001), which
sometimes allow an independent measure, or limit for the 
magnetic field strength. In general we find our equipartition 
field strength is roughly 
consistent with, or slightly lower than these estimates.

\subsection{IGM Energy Supply from Radio Sources outside of 
Rich Clusters}

It is widely believed that the energy contained in extended 
radio galaxies is supplied by the central supermassive black hole. 
The average estimated $E_{min}^{tot}$ for the ``giant'' sources reaches
$\sim 10^{61}$ ergs, which is a significant fraction of the 
gravitational energy released during the formation of 
supermassive black holes in the centers of these active galaxies
(i.e., $\sim 10^{62}$ ergs for a $10^8 M_{\odot}$ black hole).

\subsubsection{$E_{min}^{tot}$ as a minimum estimate for the total 
black hole energy}

It is important to realize that the observed magnetic energy in radio 
sources will {\it understate} the true total magnetic
energy released by the accretion onto black holes. Some additional 
energy will have been lost 
through various processes as magnetic fields are 
transported from near the black
hole (size of $\sim$ AU) to the lobes ($\sim$ Mpc). Further, after the 
energy has been deposited from the jet into the lobes, the
relativistic particles will lose energy through radiative cooling by 
(i) inverse-Compton and (ii) synchrotron radiation losses, 
as well as by possible (iii) ionization losses and (iv) particle
escape. In addition, given the very large volumes of these sources, 
some unknown amount of (v) $PdV$ work, and (vi) whatever free 
expansion energy will have been expended by the time we observe the
source. To these unquantified additions to the BH 
energy release, we must further augment the calculated minimum 
energies in Tables 1 and 2 by an increase to $V$ due to 
deprojection and also due to  
any faint undetected synchrotron halo volumes that the limited sensitivity 
of a radio map may not have revealed --- as mentioned above. 

In addition, as stated above, these minimum energies  are presumed to 
be half magnetic, the remaining half being either $49.5\% $ protons of 
any energy and $0.5\% $  electrons of gamma = $10^4$ to $10^5 $(10 to 100 
Gev), or all electrons of this energy. These energetic 
particles are accelerated by an unknown mechanism, and it is important to note that the total 
energy in a giant radio source is $\sim 10^7$  greater than the same CR's within our Galaxy. The efficiency of acceleration may be much less in many cases and thus the magnetic fields correspondingly larger.

Finally, our adopted $\nu _{2}$=$10 $GHz, hence $L$ (col. (5)) 
can be demonstrated to be conservatively low for some sources, 
which are known to radiate above $10 $GHz. Thus any 
or all of the foregoing energy losses and systematic energy 
underestimates will not 
be represented in our $E_{min}^{tot}$ tabulated in col. (7), so that 
these values will tend to understate the true 
total energy released by the AGN black hole/accretion disk. The 
important point is that they are firm lower limits. The giant 
radio sources, as we shall demonstrate below, have accumulated 
the largest amount of cosmic ray and magnetic energy outside of 
the parent galaxy. We will argue that they are especially well suited as 
probes, or ``calorimeters'' of the accumulated black hole magnetic energy 
released into the IGM space over the source's radiative 
lifetime. Because such sources last only a fraction, $\leq 1 \%$, of 
a Hubble time, they are important fiducial systems for calculating 
the total magnetic energy of the mature Universe. 

The typical volume occupied by the lobes of a single galaxy in Table 1, 
is $\sim 10^{72-73}$ cm$^{3}$, a fraction of Mpc$^3$. 
Even at these large volumes, these lobes appear to be over-pressured 
compared to the surrounding medium. This is indicated by
our estimates of the mean minimum magnetic field 
strengths, $\sim 5 \mu G$, corresponding to magnetic pressures 
$\sim 10^{-12}$ dyn cm$^{-2}$ that are much higher than the
typical thermal pressure of the IGM --- $\sim 10^{-15} n_{-5} T_6$ 
dyn cm$^{-2}$ for an assumed mean IGM density of $10^{-5}$ 
cm$^{-3}$ and a temperature of $10^6$ K. These giant radio lobes 
are therefore expected to expand further, occupying even larger
 volumes as they evolve. The details of
this volume filling process are yet to be understood.

\subsubsection{Magnetic energy as ``captured'' energy release from 
galactic black holes}

It follows from the above discussion that a substantial fraction of the energy
stored in extended extragalactic radio sources is probably in the 
form of magnetic energy. This presents a quite different picture 
from other forms of energy release, such as the intense radiation 
from AGNs. Apart from ionizing the medium, the radiation energy 
quickly loses its dynamical impact when the surrounding medium 
becomes optically thin. For magnetic fields, however,
most of its energy has been retained/confined within a large
volume (large compared to its ``engine'' size but much smaller than
the volume radiation would have filled)
for a significant fraction of cosmic time.
An important consequence of outward transported magnetic fields is that
this energy remains {\em dynamically} important perhaps for
the age of the Universe, thereby providing a {\it much stronger 
interaction with the surrounding matter} than the radiation will have.

\subsection{Comparison of the Energy Content of ``Giant'' and 
``Cluster'' Sources}

Figure 1, in which we separate the ``giant'' and ``cluster'' sample, 
shows the interrelations between the quantities tabulated above. 
The plot of size LLS vs. luminosity $L$ in Fig. 1(a) 
shows that the average radio 
luminosity for cluster radio sources is generally lower than that of
the ``giant'' sources, although there is considerable overlap. 
Although the non-cluster sources do not include those with 
LLS $<\, 0.67\,h_{75}^{-1}$ Mpc, and are not complete to a fixed lower flux 
density or luminosity
limit, they are consistent with well-established monochromatic radio 
power ($P$) vs. 
LLS plots, that also show little correlation of $P$ with LLS 
over a large range of source size (e.g. Cotter et al. 1996). 

By contrast, when we plot $E_{min}^{tot}$ against
LLS or luminosity or volume (panels (b), (c) and (d), respectively), 
a striking separation occurs between sources in these 
two different environments. The mean energy and volume 
of cluster sources are smaller by a factor of $\sim 100$ to $1000$. 
Since there is considerable overlap in 
both luminosity and redshift for these two groups, these striking 
differences in size and energy (which partially depend on each other) 
cannot be explained by any luminosity or redshift 
selection effect. Since we are observing an ensemble of each type 
of radio source 
over different evolutionary stages, and since $E_{\rm B}$ will 
gradually build up over time, the {\it upper envelope} in 
$E_{min}^{tot}$ (and $E_{\rm B}$) is significant in the context of 
this comparison. For example, in a given ICM or IGM environment, 
$E_{\rm B}(t)$ presumably builds up as the source lobes grow in 
volume, i.e. we would expect some evolutionary migration
upwards and to the right in the $E_{\rm B}$ vs. volume plot in Fig. 1. 

It will be important to understand the cause for this large 
difference in total minimum energy between cluster and giant sources. 
We can hypothesize that the typical black hole masses of cluster 
radio galaxies are smaller than the giant sources, so that they 
inherently release less magnetic energy. 
Alternatively, are AGNs in clusters activated via a different path than 
those in the typical IGM? Is magnetic field energy dissipated 
in a different way when the external plasma pressure is different? 
Unfortunately we do not understand these systems well enough to
completely rule out some of these possibilities. 

If, on the other hand, we make the reasonable assumption that 
the magnetic energy produced by the central
black holes is of the same order for both giant and cluster sources,
then most ($\sim 99\%$) of the cluster source energy has been lost 
to the ICM, so that the inferred $E_{min}^{tot}$ of cluster 
core-embedded sources now is only
a tiny residue. Note that the absolute magnitude of this energy is 
dependent upon various parameters (e.g. the CR proton component) 
and the minimum energy assumption.
The latest generation of cluster X-ray images enable an 
{\em independent} energy calculation for the cluster core sources. 
For at least two sources, Hydra A and Perseus A, the inferred $PdV$ 
work done to produce the observed X-ray ``holes'' that coincide
 with the radio lobes is 

\begin{equation}
E_{pdV} \approx n kT\,dV\,\simeq \,10^{59}n_{-3}T_{8}V_{70}\,{\rm ergs} 
\end{equation} 
where $n_{-3}$ denotes normalization to $10^{-3}{\rm cm}^{-3}$,
$T_{8}$ to $10^{8}{\rm K}$, and $V_{70}$ to $10^{70}\,{\rm cm}^{3}$,
which corresponds to a sphere with a radius of $45$ kpc. 
We assume here that the pressure and temperature are constant
over this region since the dimensions of the hot gas voids are
small compared to the cluster cores. We find that $E_{pdV}$ is roughly
{\it the same} as the estimated $E_{min}^{tot}$ of cluster sources.
This implies that the magnetic energy output from the central 
AGNs in clusters is larger than $E_{pdV}$ by at most a factor of a few. 
This would still leave their estimated total energy release 
significantly smaller than the estimated $E_{min}^{tot}$ for giant sources,
even with $k=0$, which reduces the giant source energy by approximately a factor
of 10 to $\sim 10^{59-60}$ ergs. Furthermore, we might reasonably 
expect that lobes in the IGM have experienced an equal 
or larger $PdV$ work than the ICM sources, due to expansion into a much 
larger volume (by a factor of $10^3$), even though the thermal 
pressure of the ICM is higher than the IGM by a factor of $\sim 10^{3-4}$.

Apart from this unresolved energy difference between
giant and cluster sources, they nevertheless have injected
an enormous amount of magnetic energy into their environments.
We now examine this aspect in more detail.

\section{THE IMPACT OF MAGNETIC FIELDS}
\label{sec:impact}

\subsection{Magnetic Fields in Intracluster Medium}

As discussed in the Introduction, there is now ample evidence 
that large volumes of the ICM are magnetized, with a total 
magnetic field energy
$> 10^{61}$ ergs within the central 500 kpc region of normal rich 
galaxy clusters (Clarke, Kronberg, \& B\"{o}hringer 2001). It has 
been argued (Colgate \& Li 1999; 2000) that AGNs are responsible 
for the magnetization of the ICM, motivated by the enormous 
magnetic flux as well as the
large magnetic energy in the ICM. One central point that was emphasized
in Colgate \& Li (2000) is that black hole accretion disks, 
besides being responsible for
the magnetic energy, may be the most effective magnetic flux 
multiplier or dynamo in the Universe.
A total net flux of $\sim 8\times 10^4 \mu G$ kpc$^2$ 
can be inferred from the characteristic size scale of $\sim 50$ kpc in, 
for example, Hydra A (Taylor \& Perley 1993; Colgate \& Li 2000).

Depending on the typical total magnetic energy released by a
single AGN, a total of $10^{61}/10^{59-60} \sim 10-100$ AGNs are
needed in the lifetime of a cluster in order to fill the cluster
with the measured field strength and flux, i.e. 
it is quite reasonable for AGNs to supply virtually all 
the magnetic energy in a cluster, without the need for
other processes such as a turbulent dynamo in the ICM. This is 
supported by the recently discovered fact that intracluster field
strengths out to $r \sim 500$kpc are a significant fraction of 
those found in the inner $r \leq 100$kpc core zones (Clarke et al. 2001). 
This would considerably relax the requirement for magnetic field 
amplification by turbulence in the inner, high density zones 
($r \leq 100$kpc) of cooling flow clusters. Alternatively stated, 
if cooling flow-related turbulence is a later stage of ICM 
evolution, the pre-cooling flow fields are already nearly as 
strong (Clarke et al. 2001).

Another important effect from these magnetic fields is the heating of 
the ICM due to magnetic energy dissipation. The addition of a comparable 
energy component in magnetic fields to the total thermal energy in 
the ICM could potentially change our understanding of the cluster 
structure and energetics in a fundamental way. 

\subsection{Magnetic Fields in the Wider Intergalactic Medium}

It has been suggested earlier that the intergalactic medium can be
permeated by the magnetic fields from star-driven, 
magnetized galactic winds at very early epochs, both before 
$z \sim 6$ (cf. Kronberg, Lesch, \& Hopp 1999) and since the 
clusters formed (cf. V\"{o}lk, \& Atoyan 2000). 
The above discussion has focused on an additional, and potentially
more energetic route for the IGM magnetization.
Just as AGNs can magnetize the ICM, radio-loud AGNs outside 
of clusters can be responsible for the magnetic fields in the wider IGM. 
An interesting question we may ask ourselves is 
``What happens to those giant radio lobes when the central 
AGN activities have ceased?'' We now examine this question.

To estimate the total magnetic energy generated by radio-loud AGNs,
we make the assumption that individual radio-loud QSOs (RLQSOs) 
will produce roughly similar magnetic energies as radio galaxies,
which are the sources studied here. 
This assumption is supported by the observation that
extended RLQSOs have globally similar radio properties to 
radio galaxies at the same cosmological epoch. 
The assumption is important because high-redshift 
RLQSOs may be far more abundant than 
high-redshift radio galaxies, with the result that their contribution to 
magnetic fields in the IGM begins with the cosmic epoch at which a significant
co-moving density of quasars is built up.

We estimate the mean magnetic energy density from RLQSOs as follows: 
The present total black hole mass density based on QSOs is

\begin{equation}
\rho_{\rm BH} \geq 2.2\times 10^5 / \epsilon_{0.1}~{\rm M}_{\odot}
~{\rm Mpc}^{-3} 
\end{equation}
where $\epsilon_{0.1}$ indicates that the efficiency for generating
radiation is $0.1$ 
(cf. Soltan 1982; Chokshi \& Turner 1992; Small \& Blandford 1992).
Depending on the QSO luminosity function evolution, about half
of this mass density is already accumulated by $z \approx 2$
(cf. Figure 1 of Chokshi \& Turner 1992). If we assume that
only $10\%$ of all QSOs are radio-loud (i.e., make powerful radio 
jets), and for those RLQSOs, about $10\%$ of the black hole 
accretion energy is converted into magnetic fields, then
the mean IGM {\it magnetic field} energy density by $z \approx 2$ is

\begin{equation}
e_{\rm B} \approx 5\times 10^{-3} 
\left(\frac{\epsilon_{\rm RL}}{0.1}\right)
\left(\frac{\epsilon_{\rm B}}{0.1}\right)
\rho_{\rm BH} \approx 7.3 \times 10^{-17}~{\rm ergs/cm}^{3}~,
\end{equation}
where $\epsilon_{\rm RL}$ and $\epsilon_{\rm B}$ stand for
the fraction of RLQSOs of all QSOs 
and the efficiency for them making magnetic
fields, respectively. This energy density is comparable to 
the thermal pressure of the IGM at $z \approx 2$, 
\begin{equation}
p_{\rm IGM} \approx 1.6\times 10^{-16}
\left(\frac{n_{-4}}{10^{-4}}\right)
\left(\frac{T_{4}}{10^{4}}\right)~ {\rm ergs/cm}^{-3}~. 
\end{equation}
In other words, if all the magnetic field energy were spread out
throughout the whole universe, the IGM would be a 
$\beta = p_{\rm IGM}/e_B \approx 2$ 
plasma, {\it with a comparable thermal and magnetic pressure}.

The actual impact of these magnetic fields on the IGM will
be determined by whether these magnetic fields can indeed fill up
the whole (or a significant fraction of the) volume of the IGM. 
Note that the visible magnetized lobes are created in a very short time 
($\sim 10^{7-8}$ yrs) compared to the age of the universe.
Since they could be over-pressured relative to the surrounding medium 
by a large factor (at least $\sim 100$), further expansion seems 
inevitable.  Estimates by  Furlanetto \& Loeb (2001) 
suggest that the magnetic fields
from QSOs can fill up $5-20\%$ of the IGM volume, comparable to 
Kronberg, Lesch \& Hopp's (1999) estimate of the starburst-driven 
IGM filling that is most effective at $z \geq 7$. Since magnetic 
fields made by AGNs are likely to be highly 
structured, we expect that their expansion might be quite
different from the usual adiabatic expansion. However, detailed
calculations are needed to show this.

The physics of this expansion holds the key to
a {\em quantitative} understanding of the impact of magnetic
fields on the dynamics of the IGM. Since the QSO activity peaks 
around $z \sim 2-3$, the subsequent baryonic dynamics, and the 
formation of galaxies and of large scale structure approaching galaxy scales are likely to be modified significantly by these magnetic fields.

A highly magnetized IGM appears tentatively consistent with the 
discovery of diffuse, 326 MHz synchrotron 
emission well beyond the boundaries of the Coma cluster of galaxies 
by Kim et al. (1989). The low-level synchrotron ``glow'' that
they found extending beyond the Coma cluster gave supra-cluster 
intergalactic $B_{minE}$ values between $10^{-7}$ and $10^{-6}$G over 
linear dimensions a few times the core size of the Coma cluster
itself. A consequence of our suggestion that the IGM is 
directly energized by AGN-generated magnetic flux is that more 
widespread synchrotron glow will seen be over larger IGM volumes,
which can be tested in future when 
more sensitive low frequency radio images are available.

\section{CONCLUSIONS}
\label{sec:diss}

We have analyzed the minimum energy content of the radio lobes 
of $\sim 100$ powerful radio galaxies, $\sim 70$ of which reside
in a typical low density IGM and $\sim 30$ within the inner cores of 
rich galaxy clusters.
These two groups show a large difference in the estimated
total magnetic energy, with cluster sources having 
$\sim 10^{58-59}$ ergs whereas giant sources 
have $\sim 10^{60-61}$ ergs. The latter is a significant fraction
of the total energy released from the formation of a typical
$10^{8} M_{\odot}$ black hole.

We emphasize that the observed magnetic energy understates that
made by the black hole accretion,
due to various losses incurred while magnetic fields
expand to form the giant radio lobes. This is especially
true for cluster sources where we find that a comparable or 
perhaps even larger amount of energy is expended as $PdV$ work 
in displacing the hot, dense ICM gas surrounding the lobes.

The storing of large amounts of energy in magnetic fields is a unique
way for AGNs to impact their surrounding medium. 
The AGN energy released via radiation loses its
dynamic impact when the medium becomes optically thin. 
By contrast, the AGN energy released via magnetic fields
can maintain its {\em dynamical} impact over the age of the Universe,
because most of this energy is contained in a large
volume around the galaxy. This fact may have
important consequences for galaxy and structure evolution. 

From our estimated magnetic energies arising from radio galaxies
in clusters, we argue that these AGNs can be solely responsible
for the large magnetic energy and flux, i.e., the magnetization
of the whole ICM, as revealed by recent radio and X-ray observations. 
The magnetic fields from these AGNs may also provide an important 
heating source for the whole ICM.

We further suggest that the total magnetic energy from
radio-loud QSOs/AGNs is {\em energetically} important, especially
at the epoch of $z \sim 2-3$ when QSO activity peaks. 
Giant lobes from each ``magnetic'' AGN are usually highly 
over-pressured compared to the typical IGM, thus further expansion 
of these lobes (after the central AGN activity has ceased) 
is likely to provide the space-volume filling process that could
magnetize the whole (or a significant fraction of the) IGM.
Detailed calculations of such processes will be presented 
in future publications.

\section{Acknowledgements}We acknowledge useful conversations
with T. Able, R. Daly, and M. Norman at an Aspen winter workshop
on Galaxy Formation.
P.P.K. and Q.W.D. acknowledge support from an NSERC of Canada grant 
A5713, and PPK is grateful for support from the Canada Council's 
Killam Program. The research of H.L. and S.A.C. is performed under 
the auspices of the U.S. Department of Energy, and is supported in 
part by an IGPP/Los Alamos grant and the Laboratory Directed 
Research and Development Program 
at Los Alamos.

\begin{deluxetable}{lcccccccc}
\small
\tablenum{1}
\tablecaption{Giant radio source properties}
\tablehead{
\colhead{Source} & \colhead{Name} & \colhead{z} & \colhead{$\alpha$} & \colhead{Luminosity} & \colhead{Volume} & \colhead{${\rm E_{min}(tot)}$} & \colhead{${\rm B_{min}(tot)}$} & \colhead{Refs} \\
\colhead{} & \colhead{}  & \colhead{} & \colhead{} &\colhead{${\rm(ergs/s)}$} & \colhead{${\rm(cm^3)}$} & \colhead{(ergs)} & \colhead{(gauss)} & \colhead{}
}
\startdata
0017-205 & MRC & 0.197 & -0.78 & $4.85 \times 10^{42}$ & $5.02 \times 10^{71}$ & $8.34 \times 10^{59}$ & $1.37 \times 10^{-5}$ & 1 \\
0050+402 & \nodata & 0.1488 & -0.82 & $1.56 \times 10^{42}$ & $1.45 \times 10^{72}$ & $6.95 \times 10^{59}$ & $7.35 \times 10^{-6}$ & 2,3 \\
0055+300 & NGC315 & 0.0167 & -0.59 & $1.27 \times 10^{41}$ & $2.68 \times 10^{72}$ & $2.18 \times 10^{59}$ & $3.03 \times 10^{-6}$ & 4,5 \\
0109+492 & 3C35 & 0.067 & -0.86 & $1.71 \times 10^{42}$ & $8.59 \times 10^{71}$ & $5.90 \times 10^{59}$ & $8.81 \times 10^{-6}$ & 6 \\
0114-476 & PKS & 0.146 & -0.47 & $1.01 \times 10^{43}$ & $6.07 \times 10^{72}$ & $3.62 \times 10^{60}$ & $8.22 \times 10^{-6}$ & 7 \\
0132+376 & 3C46 & 0.4373 & -1.03 & $4.87 \times 10^{43}$ & $2.46 \times 10^{71}$ & $2.36 \times 10^{60}$ & $3.30 \times 10^{-5}$ & 8 \\
0136+397 & 4C39.04 & 0.2107 & -0.87 & $6.59 \times 10^{42}$ & $1.02 \times 10^{72}$ & $1.36 \times 10^{60}$ & $1.23 \times 10^{-5}$ & 8,9 \\
0157+405 & 4C40.08 & 0.078 & -0.92 & $1.33 \times 10^{42}$ & $4.06 \times 10^{72}$ & $9.98 \times 10^{59}$ & $5.27 \times 10^{-6}$ & 10 \\
0211+326 & \nodata & 0.2605 & -0.84 & $5.92 \times 10^{42}$ & $3.78 \times 10^{72}$ & $2.23 \times 10^{60}$ & $8.16 \times 10^{-6}$ & 2,3 \\
0211-479 & PKS & 0.2195 & -0.83 & $1.04 \times 10^{43}$ & $8.22 \times 10^{71}$ & $1.60 \times 10^{60}$ & $1.48 \times 10^{-5}$ & 7 \\
0309+411 & B3 & 0.136 & -0.8 & $1.41 \times 10^{42}$ & $8.46 \times 10^{72}$ & $1.39 \times 10^{60}$ & $4.32 \times 10^{-6}$ & 11 \\
0313+683 & WENSS & 0.0902 & -0.95 & $1.22 \times 10^{42}$ & $2.86 \times 10^{72}$ & $8.15 \times 10^{59}$ & $5.68 \times 10^{-6}$ & 12 \\
0313-271 & MRC & 0.216 & -1.14 & $4.99 \times 10^{42}$ & $7.72 \times 10^{71}$ & $1.06 \times 10^{60}$ & $1.25 \times 10^{-5}$ & 1 \\
0319-454 & PKS & 0.0633 & -0.75 & $2.40 \times 10^{42}$ & $5.05 \times 10^{72}$ & $1.52 \times 10^{60}$ & $5.84 \times 10^{-6}$ & 13,14 \\
0424-728 & PKS & 0.1921 & -1.05 & $7.79 \times 10^{42}$ & $1.13 \times 10^{72}$ & $1.59 \times 10^{60}$ & $1.26 \times 10^{-5}$ & 7 \\
0437-244 & MRC & 0.84 & -0.94 & $7.61 \times 10^{43}$ & $1.81 \times 10^{71}$ & $2.61 \times 10^{60}$ & $4.03 \times 10^{-5}$ & \nodata \\
0448+519 & 3C130 & 0.109 & -0.85 & $5.60 \times 10^{42}$ & $8.93 \times 10^{71}$ & $1.18 \times 10^{60}$ & $1.22 \times 10^{-5}$ & 4 \\
0503-286 & MRC & 0.038 & -1.1 & $7.35 \times 10^{41}$ & $1.97 \times 10^{72}$ & $5.23 \times 10^{59}$ & $5.48 \times 10^{-6}$ & 15,16 \\
0511-305 & PMN & 0.0583 & -0.84 & $1.73 \times 10^{42}$ & $4.72 \times 10^{71}$ & $4.59 \times 10^{59}$ & $1.05 \times 10^{-5}$ & 7 \\
0634-205 & PMN & 0.056 & -0.87 & $4.13 \times 10^{42}$ & $3.31 \times 10^{71}$ & $6.50 \times 10^{59}$ & $1.49 \times 10^{-5}$ & 17,18 \\
0648+733 & \nodata & 0.1145 & -0.66 & $1.84 \times 10^{42}$ & $1.93 \times 10^{72}$ & $8.55 \times 10^{59}$ & $7.08 \times 10^{-6}$ & 2,3 \\
0654+482 & 7C & 0.776 & -0.75 & $1.33 \times 10^{43}$ & $3.31 \times 10^{72}$ & $3.14 \times 10^{60}$ & $1.04 \times 10^{-5}$ & 3,19 \\
0658+490 & \nodata & 0.065 & -0.64 & $2.60 \times 10^{41}$ & $6.04 \times 10^{71}$ & $1.71 \times 10^{59}$ & $5.66 \times 10^{-6}$ & 2,3 \\
0707-359 & PKS & 0.2182 & -0.72 & $1.56 \times 10^{43}$ & $2.19 \times 10^{72}$ & $3.04 \times 10^{60}$ & $1.25 \times 10^{-5}$ & 7 \\
0744+558 & DA240 & 0.0356 & -0.89 & $1.13 \times 10^{42}$ & $6.45 \times 10^{72}$ & $1.11 \times 10^{60}$ & $4.41 \times 10^{-6}$ & 5,20 \\
0813+758 & \nodata & 0.2324 & -0.74 & $6.57 \times 10^{42}$ & $6.39 \times 10^{72}$ & $2.93 \times 10^{60}$ & $7.20 \times 10^{-6}$ & 2,3 \\
0821+695 & 8C & 0.538 & -1.14 & $9.51 \times 10^{42}$ & $2.38 \times 10^{72}$ & $2.53 \times 10^{60}$ & $1.10 \times 10^{-5}$ & 21,22 \\
0915+320 & B2 & 0.062 & -0.5 & $1.64 \times 10^{41}$ & $1.08 \times 10^{71}$ & $6.29 \times 10^{58}$ & $8.10 \times 10^{-6}$ & 23 \\
0945+734 & 4C73.08 & 0.0581 & -0.85 & $1.44 \times 10^{42}$ & $1.58 \times 10^{72}$ & $6.94 \times 10^{59}$ & $7.05 \times 10^{-6}$ & 24 \\
1003+351 & 3C236 & 0.0988 & -0.61 & $7.25 \times 10^{42}$ & $1.01 \times 10^{73}$ & $3.81 \times 10^{60}$ & $6.53 \times 10^{-6}$ & 5 \\
1025-229 & MRC & 0.309 & -0.9 & $9.95 \times 10^{42}$ & $2.00 \times 10^{71}$ & $8.54 \times 10^{59}$ & $2.20 \times 10^{-5}$ & \nodata \\
1029+571 & HB13 & 0.034 & -0.85 & $1.13 \times 10^{41}$ & $6.91 \times 10^{70}$ & $4.25 \times 10^{58}$ & $8.34 \times 10^{-6}$ & 4,25 \\
1058+368 & 7C & 0.75 & -0.75 & $1.98 \times 10^{43}$ & $3.24 \times 10^{72}$ & $3.92 \times 10^{60}$ & $1.17 \times 10^{-5}$ & 3,19 \\
1127-130 & PKS & 0.6337 & -0.87 & $7.28 \times 10^{43}$ & $1.60 \times 10^{71}$ & $2.37 \times 10^{60}$ & $4.10 \times 10^{-5}$ & 26 \\
1144+352 & WENSS & 0.063 & -0.56 & $3.78 \times 10^{40}$ & $2.10 \times 10^{71}$ & $3.61 \times 10^{58}$ & $4.42 \times 10^{-6}$ & 27,28 \\
1158+351 & 87GB & 0.55 & -1.1 & $1.90 \times 10^{43}$ & $4.99 \times 10^{71}$ & $1.91 \times 10^{60}$ & $2.08 \times 10^{-5}$ & 29 \\
1209+745 & 4C74.17 & 0.107 & -0.85 & $1.13 \times 10^{42}$ & $4.77 \times 10^{71}$ & $3.60 \times 10^{59}$ & $9.24 \times 10^{-6}$ & 30 \\
1213+422 & \nodata & 0.2426 & -0.83 & $3.66 \times 10^{42}$ & $4.06 \times 10^{72}$ & $1.74 \times 10^{60}$ & $6.97 \times 10^{-6}$ & 2,3 \\
1218+639 & TXS & 0.2 & -0.85 & $2.62 \times 10^{42}$ & $3.11 \times 10^{72}$ & $1.29 \times 10^{60}$ & $6.86 \times 10^{-6}$ & 31 \\
1232+216 & 3C274.1 & 0.422 & -0.92 & $7.83 \times 10^{43}$ & $4.32 \times 10^{71}$ & $3.86 \times 10^{60}$ & $3.18 \times 10^{-5}$ & 32 \\
1309+412 & \nodata & 0.1103 & -0.83 & $9.51 \times 10^{41}$ & $3.66 \times 10^{71}$ & $2.91 \times 10^{59}$ & $9.49 \times 10^{-6}$ & 2,10 \\
1312+698 & DA340 & 0.106 & -0.73 & $2.48 \times 10^{42}$ & $5.56 \times 10^{71}$ & $5.99 \times 10^{59}$ & $1.10 \times 10^{-5}$ & 2,31 \\
1331-099 & PKS & 0.081 & -0.9 & $2.47 \times 10^{42}$ & $1.42 \times 10^{72}$ & $9.04 \times 10^{59}$ & $8.48 \times 10^{-6}$ & 3,33 \\
1349+647 & 3C292 & 0.71 & -0.8 & $2.08 \times 10^{44}$ & $9.13 \times 10^{70}$ & $3.32 \times 10^{60}$ & $6.41 \times 10^{-5}$ & 34,35 \\
1358+305 & B2 & 0.206 & -0.99 & $3.77 \times 10^{42}$ & $5.83 \times 10^{72}$ & $2.11 \times 10^{60}$ & $6.40 \times 10^{-6}$ & 36 \\
1426+295 & \nodata & 0.087 & -0.78 & $4.34 \times 10^{41}$ & $1.40 \times 10^{72}$ & $3.30 \times 10^{59}$ & $5.17 \times 10^{-6}$ & 2,3 \\
1450+333 & \nodata & 0.249 & -0.94 & $4.57 \times 10^{42}$ & $3.79 \times 10^{72}$ & $1.95 \times 10^{60}$ & $7.63 \times 10^{-6}$ & 2,3 \\
1452-517 & MRC & 0.08 & -0.3 & $3.45 \times 10^{42}$ & $4.03 \times 10^{72}$ & $1.66 \times 10^{60}$ & $6.83 \times 10^{-6}$ & 37 \\
1519+513 & 87GB & 0.37 & -0.88 & $3.29 \times 10^{43}$ & $1.08 \times 10^{72}$ & $3.48 \times 10^{60}$ & $1.90 \times 10^{-5}$ & 29 \\
1543+845 & \nodata & 0.201 & -0.89 & $2.26 \times 10^{42}$ & $3.15 \times 10^{72}$ & $1.20 \times 10^{60}$ & $6.56 \times 10^{-6}$ & 2,3 \\
1545-321 & PKS & 0.1085 & -0.94 & $3.43 \times 10^{42}$ & $4.71 \times 10^{71}$ & $6.80 \times 10^{59}$ & $1.28 \times 10^{-5}$ & 7 \\
1549+202 & 3C326 & 0.0895 & -0.82 & $4.79 \times 10^{42}$ & $5.02 \times 10^{72}$ & $2.26 \times 10^{60}$ & $7.13 \times 10^{-6}$ & 5,38 \\
1602+376 & 7C & 0.814 & -0.75 & $2.28 \times 10^{43}$ & $5.83 \times 10^{72}$ & $5.43 \times 10^{60}$ & $1.03 \times 10^{-5}$ & 3,19 \\
1626+518 & WENSS & 0.056 & -0.66 & $3.31 \times 10^{41}$ & $4.14 \times 10^{71}$ & $1.68 \times 10^{59}$ & $6.77 \times 10^{-6}$ & 2,39 \\
1636+418 & 7C & 0.867 & -0.75 & $1.24 \times 10^{43}$ & $2.76 \times 10^{72}$ & $2.77 \times 10^{60}$ & $1.07 \times 10^{-5}$ & 3,19 \\
1637+826 & NGC6251 & 0.023 & -0.58 & $3.68 \times 10^{41}$ & $2.94 \times 10^{72}$ & $4.15 \times 10^{59}$ & $4.00 \times 10^{-6}$ & 5,40 \\
1701+423 & 7C & 0.476 & -0.75 & $8.89 \times 10^{42}$ & $3.60 \times 10^{72}$ & $2.66 \times 10^{60}$ & $9.13 \times 10^{-6}$ & 3,19 \\
1721+343 & 4C34.47 & 0.2055 & -0.75 & $9.44 \times 10^{42}$ & $5.28 \times 10^{71}$ & $1.24 \times 10^{60}$ & $1.63 \times 10^{-5}$ & 41 \\
1834+620 & WENSS & 0.519 & -0.97 & $4.23 \times 10^{43}$ & $2.14 \times 10^{71}$ & $2.03 \times 10^{60}$ & $3.28 \times 10^{-5}$ & 42,43 \\
1910-800 & PKS & 0.346 & -0.91 & $1.88 \times 10^{43}$ & $1.08 \times 10^{72}$ & $2.53 \times 10^{60}$ & $1.63 \times 10^{-5}$ & 7 \\
1918+516 & \nodata & 0.284 & -0.91 & $4.87 \times 10^{42}$ & $5.49 \times 10^{72}$ & $2.36 \times 10^{60}$ & $6.97 \times 10^{-6}$ & 2,3 \\
2043+749 & 4C74.26 & 0.104 & -0.81 & $2.53 \times 10^{42}$ & $1.89 \times 10^{72}$ & $1.03 \times 10^{60}$ & $7.84 \times 10^{-6}$ & 2,44 \\
2147+816 & \nodata & 0.1457 & -0.45 & $1.89 \times 10^{42}$ & $5.75 \times 10^{72}$ & $1.36 \times 10^{60}$ & $5.17 \times 10^{-6}$ & 2,45 \\
2309+184 & 3C457 & 0.427 & -1.02 & $8.08 \times 10^{43}$ & $4.74 \times 10^{71}$ & $4.18 \times 10^{60}$ & $3.16 \times 10^{-5}$ & 46 \\
\enddata
\tablerefs{1. Kapahi et al. (1998); 2. Schoenmakers et al. (2000a); 3. Condon et al (1998); 
4. Jagers (1987b); 5. Mack et al. (1997); 6. Jagers (1987a); 
7. Subrahmanyan, Saripalli \& Hunstead (1996); 8. Gregorini et al. (1988); 
9. Hine (1979); 10. Vigotti et al. (1989); 11. de Bruyn (1989); 
12. Schoenmakers et al. (1998); 13. Jones \& McAdam (1992); 14. Saripalli, Subrahmanyan \& Hunstead (1994);
15. Saripalli et al. (1986); 16. Subrahmanya \& Hunstead (1986); 17. Kronberg, Wielebinski \& Graham (1986); 
18. Danziger et al. (1978); 19. Cotter, Rawlings \& Saunders (1996); 20. Strom, Baker \& Willis (1981); 
21. Lacy et al. (1993); 22. Lara et al. (2000); 23. Ekers et al. (1981); 24. Mayer (1979); 
25. Masson (1979); 26. Bhatnagar, Gopal-Krishna \& Wisotzki (1998); 27. Schoenmakers et al. (1999);
28. Snellen et al. (1995); 29. Machalski \& Condon (1985); 30. van Breugel \& Willis (1981);
31. Saunders, Baldwin \& Warner (1987); 32. Strom et al. (1990); 33. Saripalli et al. (1996);
34. Alexander \& Leahy (1987); 35. Leahy, Pooley \& Riley (1986); 
36. Parma et al. (1996); 37. Jones (1986); 38. Willis \& Strom (1978); 39. R\"{o}ttgering et al. (1996);
40. Willis et al. (1982); 41. J\"{a}gers et al. (1982); 42. Schoenmakers et al. (2000b); 
43. Lara et al. (1999); 44. Riley et al. (1989); 45. Palma et al. (2000); 46. Leahy \& Perley (1991)}
\end{deluxetable}

\begin{deluxetable}{lcccccccc}
\small
\tablenum{2}
\tablecaption{Cluster source properties}
\tablehead{
\colhead{Source} & \colhead{Name} & \colhead{z} & \colhead{$\alpha$} & \colhead{Luminosity} & \colhead{Volume} & \colhead{${\rm E_{min}(tot)}$} & \colhead{${\rm B_{min}(tot)}$} & \colhead{Refs} \\
\colhead{} & \colhead{} & \colhead{} & \colhead{} & \colhead{${\rm(ergs/s)}$} & \colhead{${\rm(cm^3)}$} & \colhead{(ergs)} & \colhead{(gauss)} & \colhead{}
}
\startdata
0019+230 & 4C23.01 & 0.1332 & -1.12 & $8.76 \times 10^{41}$ & $2.69 \times 10^{69}$ & $3.44 \times 10^{58}$ & $3.81 \times 10^{-5}$ & 1 \\
0037+209 & \nodata & 0.0579 & -0.78 & $7.32 \times 10^{40}$ & $8.86 \times 10^{69}$ & $1.37 \times 10^{58}$ & $1.32 \times 10^{-5}$ & 1 \\
0043+201 & 4C20.04 & 0.1063 & -0.8 & $1.62 \times 10^{42}$ & $9.23 \times 10^{69}$ & $8.15 \times 10^{58}$ & $3.16 \times 10^{-5}$ & 2 \\
0053-015 & \nodata & 0.03822 & -1.01 & $4.35 \times 10^{41}$ & $5.55 \times 10^{69}$ & $3.13 \times 10^{58}$ & $2.52 \times 10^{-5}$ & 3,4 \\
0110+152 & \nodata & 0.0447 & -0.85 & $3.18 \times 10^{41}$ & $7.35 \times 10^{69}$ & $2.94 \times 10^{58}$ & $2.13 \times 10^{-5}$ & 2 \\
0124+189 & 4C18.06 & 0.04268 & -0.48 & $3.56 \times 10^{41}$ & $2.68 \times 10^{68}$ & $7.52 \times 10^{57}$ & $5.63 \times 10^{-5}$ & 1 \\
0154+320 & \nodata & 0.0891 & -0.88 & $4.96 \times 10^{41}$ & $1.21 \times 10^{69}$ & $1.74 \times 10^{58}$ & $4.04 \times 10^{-5}$ & 1 \\
0255+058 & 3C75 & 0.023153 & -0.78 & $4.57 \times 10^{41}$ & $2.30 \times 10^{69}$ & $2.20 \times 10^{58}$ & $3.29 \times 10^{-5}$ & 5 \\
0320-373 & Fornax A & 0.00587 & -0.55 & $6.01 \times 10^{41}$ & $1.58 \times 10^{71}$ & $1.58 \times 10^{59}$ & $1.06 \times 10^{-5}$ & 6 \\
0719+670 & 4C67.13 & 0.08723 & -0.7 & $7.08 \times 10^{41}$ & $9.02 \times 10^{68}$ & $1.87 \times 10^{58}$ & $4.84 \times 10^{-5}$ & 1 \\
0756+272 & \nodata & 0.0991 & -1.01 & $9.53 \times 10^{41}$ & $8.53 \times 10^{69}$ & $5.88 \times 10^{58}$ & $2.79 \times 10^{-5}$ & 1 \\
0803-008 & 3C193 & 0.0891 & -0.8 & $1.63 \times 10^{42}$ & $5.34 \times 10^{69}$ & $6.46 \times 10^{58}$ & $3.70 \times 10^{-5}$ & 1 \\
0836+290 & 4C29.30 & 0.0788 & -0.85 & $8.08 \times 10^{41}$ & $2.85 \times 10^{70}$ & $8.91 \times 10^{58}$ & $1.88 \times 10^{-5}$ & 2 \\
0915-118 & Hydra A & 0.053 & -0.93 & $2.20 \times 10^{43}$ & $7.57 \times 10^{67}$ & $4.65 \times 10^{58}$ & $2.64 \times 10^{-4}$ & 7 \\
1159+583 & 4C58.23 & 0.1018 & -0.8 & $1.65 \times 10^{42}$ & $2.52 \times 10^{68}$ & $1.76 \times 10^{58}$ & $8.89 \times 10^{-5}$ & 2 \\
1222+131 & 3C272.1 & 0.003429 & -0.6 & $1.03 \times 10^{40}$ & $2.19 \times 10^{66}$ & $1.28 \times 10^{56}$ & $8.13 \times 10^{-5}$ & 8,9 \\
1231+674 & 4C67.12 & 0.1062 & -0.9 & $1.75 \times 10^{42}$ & $1.75 \times 10^{69}$ & $4.20 \times 10^{58}$ & $5.21 \times 10^{-5}$ & 2 \\
1233+169 & PKS & 0.0784 & -0.51 & $5.78 \times 10^{41}$ & $6.18 \times 10^{69}$ & $3.77 \times 10^{58}$ & $2.63 \times 10^{-5}$ & 1 \\
1246-410 & NGC4696 & 0.0099 & -0.84 & $6.06 \times 10^{40}$ & $1.84 \times 10^{66}$ & $3.26 \times 10^{56}$ & $1.42 \times 10^{-4}$ & 10 \\
1256+281 & 5C4.81 & 0.0235 & -1.13 & $5.35 \times 10^{40}$ & $1.05 \times 10^{69}$ & $4.63 \times 10^{57}$ & $2.23 \times 10^{-5}$ & 11 \\
1409+52 & 3C295 & 0.461 & -0.98 & $1.08 \times 10^{45}$ & $1.08 \times 10^{67}$ & $1.87 \times 10^{59}$ & $1.40 \times 10^{-3}$ & 12 \\
1433+553 & 4C55.29 & 0.1396 & -0.7 & $1.52 \times 10^{42}$ & $8.00 \times 10^{68}$ & $2.72 \times 10^{58}$ & $6.21 \times 10^{-5}$ & 2 \\
1508+065 & \nodata & 0.08086 & -0.83 & $5.80 \times 10^{41}$ & $1.96 \times 10^{68}$ & $8.71 \times 10^{57}$ & $7.10 \times 10^{-5}$ & 1 \\
1638+558 & \nodata & 0.2426 & -0.8 & $2.75 \times 10^{42}$ & $1.14 \times 10^{70}$ & $1.19 \times 10^{59}$ & $3.43 \times 10^{-5}$ & 1 \\
1820+689 & 4C68.21 & 0.0881 & -0.63 & $9.25 \times 10^{41}$ & $7.42 \times 10^{69}$ & $5.35 \times 10^{58}$ & $2.86 \times 10^{-5}$ & 1 \\
1826+747 & \nodata & 0.121 & -0.8 & $1.41 \times 10^{42}$ & $2.62 \times 10^{68}$ & $1.63 \times 10^{58}$ & $8.40 \times 10^{-5}$ & 2 \\
1957+405 & Cygnus A & 0.056075 & -1.01 & $9.14 \times 10^{44}$ & $3.10 \times 10^{69}$ & $1.93 \times 10^{60}$ & $2.65 \times 10^{-4}$ & 13,14,15  \\
2229-086 & PKS & 0.0831 & -0.52 & $8.38 \times 10^{41}$ & $2.51 \times 10^{70}$ & $8.49 \times 10^{58}$ & $1.96 \times 10^{-5}$ & 1 \\
2236-176 & \nodata & 0.0698 & -0.55 & $1.22 \times 10^{42}$ & $8.02 \times 10^{68}$ & $2.41 \times 10^{58}$ & $5.83 \times 10^{-5}$ & 2 \\
2335+267 & 3C465 & 0.030221 & -0.84 & $1.10 \times 10^{42}$ & $5.07 \times 10^{69}$ & $5.10 \times 10^{58}$ & $3.37 \times 10^{-5}$ & 16 \\
 & Perseus A & & & & & & & \\
\raisebox{1.5ex}[0pt]{3C84} & (Halo) & \raisebox{1.5ex}[0pt]{0.017559} & \raisebox{1.5ex}[0pt]{-1.1} & \raisebox{1.5ex}[0pt]{$3.28 \times 10^{41}$} & \raisebox{1.5ex}[0pt]{$5.87 \times 10^{70}$} & \raisebox{1.5ex}[0pt]{$7.31 \times 10^{58}$} & \raisebox{1.5ex}[0pt]{$1.19 \times 10^{-5}$} & \raisebox{1.5ex}[0pt]{17,18} \\
 & Perseus A & & & & & & & \\
\raisebox{1.5ex}[0pt]{3C84} & (Inner arcmin) & \raisebox{1.5ex}[0pt]{0.017559} & \raisebox{1.5ex}[0pt]{-1.01} & \raisebox{1.5ex}[0pt]{$1.33 \times 10^{41}$} & \raisebox{1.5ex}[0pt]{$3.84 \times 10^{67}$} & \raisebox{1.5ex}[0pt]{$1.88 \times 10^{57}$} & \raisebox{1.5ex}[0pt]{$7.44 \times 10^{-5}$} & \\
 & Virgo A & & & & & & & 18,19,20 \\
\raisebox{1.5ex}[0pt]{3C274} & (Halo only) & \raisebox{1.5ex}[0pt]{0.00436} & \raisebox{1.5ex}[0pt]{-1.1} & \raisebox{1.5ex}[0pt]{$3.00 \times 10^{41}$} & \raisebox{1.5ex}[0pt]{$1.90 \times 10^{69}$} & \raisebox{1.5ex}[0pt]{$1.60 \times 10^{58}$} & \raisebox{1.5ex}[0pt]{$3.08 \times 10^{-5}$} & 21,22,23 \\
 & Virgo A & & & & & & & \\
\raisebox{1.5ex}[0pt]{3C274} & (Inner radio core) & \raisebox{1.5ex}[0pt]{0.00436} & \raisebox{1.5ex}[0pt]{-0.5} & \raisebox{1.5ex}[0pt]{$3.31 \times 10^{41}$} & \raisebox{1.5ex}[0pt]{$5.28 \times 10^{65}$} & \raisebox{1.5ex}[0pt]{$5.04 \times 10^{56}$} & \raisebox{1.5ex}[0pt]{$3.29 \times 10^{-4}$} & \\
\enddata
\tablerefs{1. Owen \& Ledlow (1997); 2. O'Donoghue et al. (1990); 3. Feretti et al. (1999); 
4. O'Dea \& Owen (1985); 
5. Owen et al. (1985); 6. Ekers et al. (1983); 
7. Taylor et al. (1990); 8. Laing \& Bridle (1987); 9. Zukowski (1990); 
10. Taylor, Allen \& Fabian (1999); 11. Dallacasa et al. (1989); 12. Perley \& Taylor (1991); 
13. Carilli et al. (1991); 14. Baars et al. (1977); 15. Hargrave \& Ryle (1974);
16. Eilek et al. (1984); 17. Pedlar et al. (1990); 18. Herbig \& Readhead (1992); 19. Kassim et al. (1993);
20. Andernach et al. (1979); 21. Rottmann et al. (1996); 22. Hines, Eilek \& Owen (1989); 23. Turland (1975)}
\end{deluxetable}

\begin{figure}
\begin{center}
\psfig{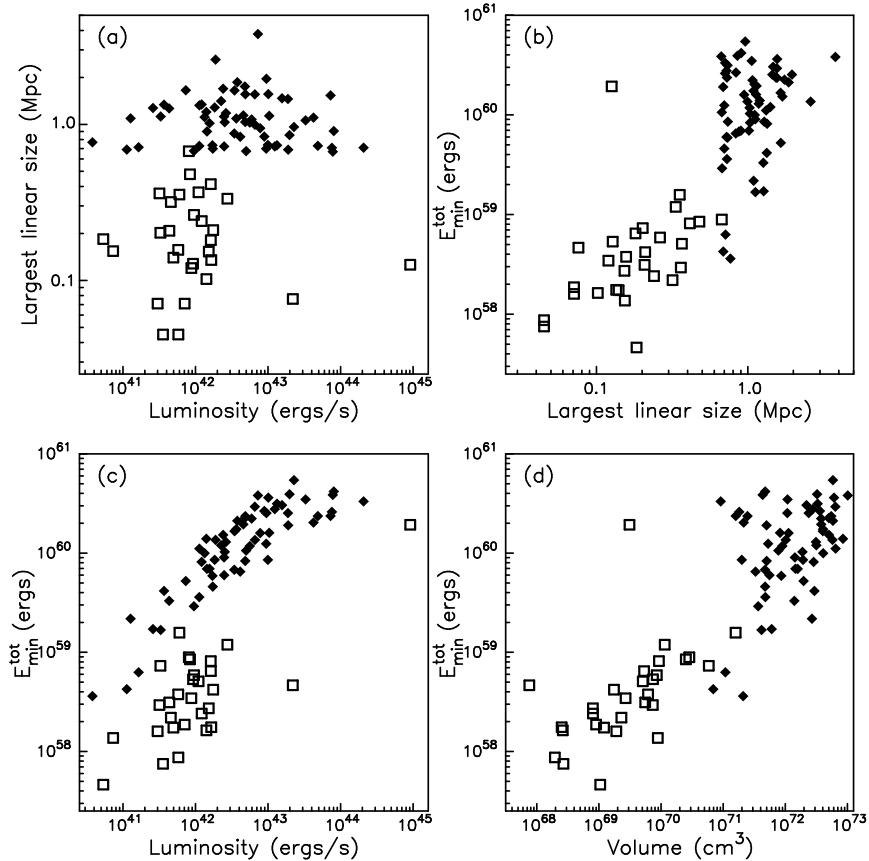}
\caption{Shown are the observed and estimated quantities
tabulated in Table 1. The giant and cluster sources are represented
by solid diamonds and open squares, respectively. Although the groups
overlap in their luminosities, their total minimum energies show
a marked difference (by a factor of $\sim 100$). A large difference,
$\sim 1000$, is seen in their estimated volumes as well. Note that three 
cluster sources, 1222+131, 1246-410, and 1409+52 are not plotted because 
of their small linear size.}
\end{center}
\end{figure}

\end{document}